\documentclass[twoside,12pt]{article}
\usepackage{epsfig}

\def\Journal#1#2#3#4{{#1} {#2} (#4) #3 }

\def\PL{{\em Phys. Lett.}}
\def\PRL{\em Phys. Rev. Lett.}

\def\PREP{\em Phys. Rep.}

\def\PRC{{\em Phys. Rev.} C}

\def\ZPA{{\em Z. Phys.} A}

\newcommand{\be}{\begin{equation}}
\newcommand{\ee}{\end{equation}}
\newcommand{\bea}{\begin{eqnarray}}
\newcommand{\eea}{\end{eqnarray}}

\begin{document}

\title{ \vspace{1cm} Differential Neutron-Proton Squeeze-out}
\author{W.\ Trautmann,$^{1}$ M.\ Chartier,$^2$ Y.\ Leifels,$^1$ 
R.C.\ Lemmon,$^{3}$ Q.\ Li,$^4$ \\ 
J.\ {\L}ukasik,$^5$ A.\ Pagano,$^6$ P.\ Paw{\l}owski,$^5$ 
P.\ Russotto,$^7$ P.\ Wu$^2$\\ 
\\
$^{1}$ GSI Darmstadt, D-64291 Darmstadt, Germany\\
$^{2}$ University of Liverpool, Liverpool L69 7ZE United Kingdom\\
$^{3}$ STFC Daresbury Laboratory, Warrington, WA4 4AD United Kingdom\\
$^{4}$ FIAS, Universit\"{a}t Frankfurt, D-60438 Frankfurt am Main, Germany\\
$^{5}$ IFJ-PAN, Pl-31342 Krak\'ow, Poland\\
$^{6}$ INFN-Sezione di Catania, I-95123 Catania, Italy\\
$^{7}$ INFN-LNS and Universit\`{a} di Catania, I-95123 Catania, Italy}

\maketitle

\begin{abstract}
The elliptic flow (squeeze-out) of neutrons, protons and light complex particles in
reactions of neutron-rich systems at relativistic energies is proposed as an 
observable sensitive to the strength of the symmetry term in the equation of state 
at supra-normal densities. Preliminary results from a study of the existing 
FOPI/LAND data for $^{197}$Au + $^{197}$Au collisions at 400 A MeV with the UrQMD 
model favor a moderately soft symmetry term with a density dependence of the
potential term proportional to $(\rho/\rho_0)^{\gamma}$ with 
$\gamma = 0.6 \pm 0.3$.

\end{abstract}
The equation of state (EOS) of asymmetric nuclear matter is of fundamental importance 
to both nuclear physics and astrophysics. However, our understanding of the EOS is 
limited, largely due to our poor knowledge of the density dependence of the nuclear 
symmetry energy. While considerable progress has been made recently in determining the 
symmetry energy around normal nuclear matter density\cite{li08,ditoro}, much more work 
is still needed to probe its high-density behaviour. This will require reaction studies
at sufficiently high energies and probes sensitive to mean-field effects in the initial
compressed stage of the reaction as, e.g., the neutron-proton differential transverse 
and elliptic flows\cite{baoan02,greco03}. 

In a series of experiments at GSI using the LAND and FOPI detectors, both neutron and 
proton collective flow observables from $^{197}$Au + $^{197}$Au collisions at 400, 600 
and 800 A MeV have been measured\cite{leifels93,lambrecht94}. This data set is presently
being reanalyzed in order to determine the optimum conditions for a dedicated 
differential 
flow experiment, but also with the aim to produce first results for a comparison with
model predictions. The theoretical analysis is based on the UrQMD model of Li et 
al. which has recently been updated for the investigation of heavy-ion 
reactions at intermediate energies\cite{qfli06}.

\begin{figure}[tb]
\begin{center}
\begin{minipage}[t]{8 cm}
\epsfig{file=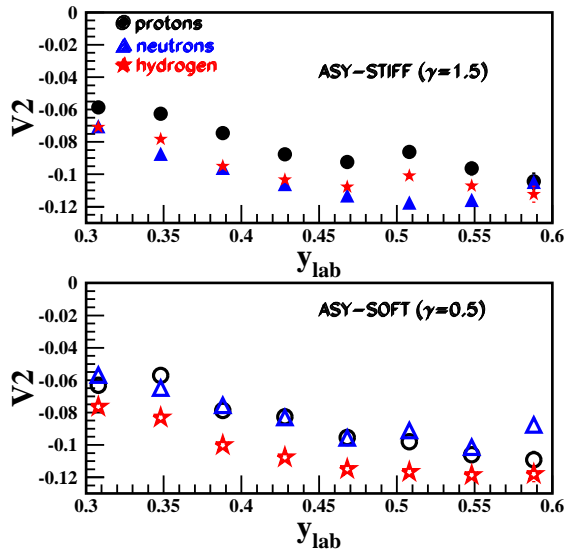,scale=0.4}
\end{minipage}
\begin{minipage}[t]{16.5 cm}
\caption{Elliptic flow parameter $v_2$ for mid-peripheral $^{197}$Au + $^{197}$Au 
collisions at 400 MeV per nucleon as calculated with the UrQMD model for protons 
(circles), neutrons (triangles), and hydrogen isotopes (stars) as a function of 
the laboratory rapidity $y_{\rm lab}$. The results have been filtered to 
correspond to the geometrical acceptance of the neutron detector used in the joint 
FOPI/LAND experiment.
\label{fig1}}
\end{minipage}
\end{center}
\end{figure}

The predictions obtained for the elliptic flow of neutrons, protons and hydrogen 
isotopes for $^{197}$Au + $^{197}$Au at 400 MeV per nucleon are shown in 
Fig.~\ref{fig1}. The asymmetry of the $v_2$ parameters with respect to midrapidity 
($y_{\rm mid} = 0.448$) is caused by the LAND acceptance which extends to larger 
transverse momenta $p_t$ at more forward rapidities\cite{leifels93}. 
The significantly larger neutron squeeze-out for a stiff density dependence 
of the symmetry term (upper panel) in comparison to the soft case (lower panel) 
is, however, nearly the same in the filtered and unfiltered observables. 


The relative strengths of the measured neutron, proton, and hydrogen elliptic flows
are found to be closer to the soft than to the stiff predictions. A linear interpolation 
between the two cases yields $\gamma = 0.6 \pm 0.3$ as a preliminary result
for the power-law exponent 
describing the density dependence of the potential term. The experimental error is,
to a large part, statistical which means that it can be improved with a new measurement.
However, data from a single reaction system will not uniquely tie the observed 
relative squeeze-out strengths  
to mean-field effects related to the neutron richness of the $^{197}$Au + $^{197}$Au 
system. For this purpose, a double-differential flow measurement for 
two systems with different $N/Z$ as, e.g., the symmetric $^{112}$Sn+$^{112}$Sn and 
$^{124}$Sn+$^{124}$Sn reactions appears mandatory\cite{yong06}.

\end{document}